\documentclass[journal,twocolumn]{IEEEtran}
\usepackage{epsfig,rotating,setspace,latexsym,amsmath,epsf,amssymb,amsfonts,bm,theorem,epstopdf}
\usepackage{cite,authblk}
\usepackage{algorithmic,algorithm}
\usepackage{setspace}
\usepackage{color}
\usepackage{graphicx}
\usepackage{mathrsfs}
\usepackage{multicol,blindtext}
\ifCLASSOPTIONcompsoc
\usepackage[caption=false, font=normalsize, labelfont=sf, textfont=sf]{subfig}
\else
\usepackage[caption=false, font=footnotesize]{subfig}
\fi

\newtheorem{lemma}{Lemma}

\newenvironment{Proof}[1]{\medskip\par\noindent{\bf Proof:\,}\,#1}{{\mbox{\,$\blacksquare$}\par}}

\allowdisplaybreaks

\begin{document}
	
\title{Age of Information for Updates with Distortion: Constant and Age-Dependent Distortion Constraints
\thanks{This work was supported by NSF Grants CNS 15-26608, CCF 17-13977 and ECCS 18-07348, and is presented in part at IEEE Information Theory Workshop, Visby, Gotland, Sweden, August 2019.}
\thanks{The authors are with the Department of Electrical and Computer Engineering, University of Maryland, College Park, MD 20742 USA (e-mail: bastopcu@umd.edu, ulukus@umd.edu).}}	
		
\author{Melih Bastopcu \qquad Sennur Ulukus}
	
\maketitle
	
\begin{abstract}	
We consider an information update system where an information receiver requests updates from an information provider in order to minimize its age of information. The updates are generated at the information provider (transmitter) as a result of completing a set of tasks such as collecting data and performing computations on them. We refer to this as the update generation process. We model the \emph{quality} of an update as an increasing function of the processing time spent while generating the update at the transmitter. In particular, we use \emph{distortion} as a proxy for \emph{quality}, and model distortion as a decreasing function of processing time. Processing longer at the transmitter results in a better quality (lower distortion) update, but it causes the update to age in the process. We determine the age-optimal policies for the update request times at the receiver and the update processing times at the transmitter subject to a minimum required quality (maximum allowed distortion) constraint on the updates. For the required quality constraint, we consider the cases of constant maximum allowed distortion constraints, as well as age-dependent maximum allowed distortion constraints. 
\end{abstract}
		
\section{Introduction}

As time-critical information is becoming ever more important, especially with the emergence of applications such as autonomous driving, augmented/virtual reality, and online gaming, a new performance metric called \emph{age of information} has been introduced to quantify the \textit{freshness} of information in communication networks. Age of information has been studied in the context of web crawling \cite{Cho03, Brewington00, Azar18, Kolobov19a}, social networks \cite{Ioannidis09}, queueing networks \cite{ Kaul12a, Costa14, Bedewy16, He16a, Kam16b, Sun17a, Najm18b, Najm17, Soysal18, Soysal19, Buyukates20e}, caching systems \cite{Gao12, Yates17b, Kam17b, Zhong18c, Zhang18, Tang19, Yang19a, bastopcu20e}, remote estimation \cite{Wang19a, Sun17b, Sun18b, Chakravorty18}, energy harvesting systems \cite{Bacinoglu15, Bacinoglu17, Bacinoglu18, Baknina18b, Baknina18a, Wu18, Feng18a, Feng18c, Arafa18a, Arafa18b, Arafa18c, Arafa18f, Arafa19e, Arafa17b, Arafa17a, Farazi18, Leng19, Chen19}, fading wireless channels \cite{Bhat19, Ostman19}, scheduling in networks \cite{Nath17, Hsu18b, Kadota18a, Kadota18b, Kosta17a, Bastopcu18, bastopcu_soft_updates_journal, Buyukates18c, Buyukates19b, Bastopcu20a, Zhong18a,Rajaraman18, Liu19}, multi-hop multicast networks \cite{ Zhong17a, Buyukates18, Buyukates19, Buyukates18b}, lossless and lossy source coding \cite{Zhong16, Zhong18f, Mayekar18, MelihBatu1, MelihBatu2, Ramirez19, Bastopcu20}, computation-intensive systems \cite{Kuang19, Gong19, Buyukates19c, Zou19b, Arafa19a, Bastopcu19, Behrouzi18}, vehicular, IoT and UAV systems \cite{ Elmagid18, Liu18, Elmagid19c, Alabbasi20}, reinforcement learning \cite{Ceran18, Beytur19, Elmagid19}, and so on.     

We consider an information update system where an information receiver requests updates from an information provider in order to minimize the age of information at the receiver. To generate an update, the information provider completes a set of tasks such as collecting data and processing them. We consider the \emph{quality} of updates via their \emph{distortion}. We model the \emph{quality} (resp., the \emph{distortion}) of an update as a monotonically increasing (resp., monotonically decreasing) function of the processing time spent to generate the update at the transmitter.

\begin{figure}[t]
	\centerline{\includegraphics[width=0.95\columnwidth]{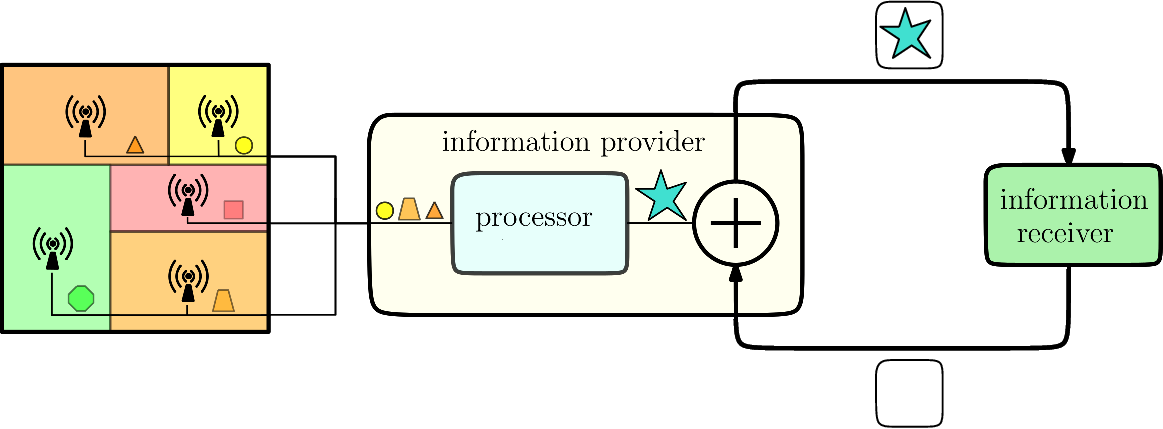}}
	\caption{An information updating system which consists of an information provider which collects/processes data and an information receiver.}
	\label{Fig:system_model}
	\vspace{-4mm}
\end{figure}

Examples of such systems can be found in sensor networking and distributed computation applications. For instance, in a sensor networking application where multiple sensors observe the realization of a common underlying random variable (e.g., temperature), if the information provider generates an update using the observation of a single sensor, the update will be generated faster, but will have large distortion; and conversely, if the information provider generates an update using the observations of all sensors, the update will be generated with a delay, but will have small distortion. Similarly, in a distributed computation system with stragglers, the master can generate an update using faster servers with lower quality, or utilize all servers to generate a better quality update with a delay. Thus, there is a trade-off between processing time and quality.

\begin{figure}[t]
	\centerline{\includegraphics[width=0.95\columnwidth]{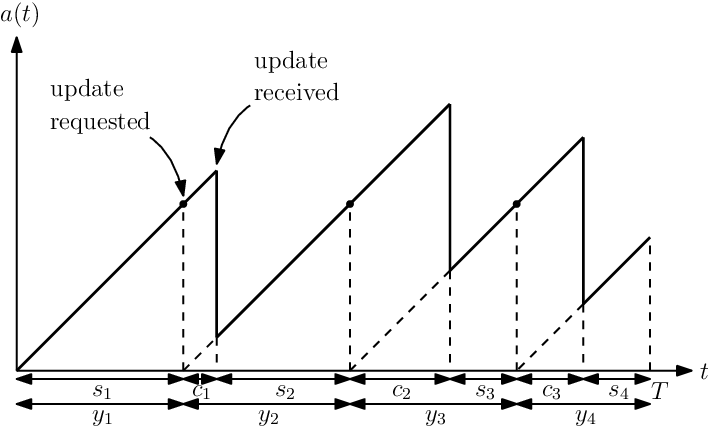}}
	\caption{Age evolution at the receiver.}
	\label{Fig:age_eval}
	\vspace{-4mm}
\end{figure}

We consider the information update system shown in Fig.~\ref{Fig:system_model}. The information provider connects to multiple units (sensors, servers, etc.) to generate an update. When there is no update, the information at the receiver gets stale over time, i.e., the age increases linearly. The information receiver requests an update from the information provider. After receiving the update request, the information provider allocates $c_i$ amount of time as shown in Fig.~\ref{Fig:age_eval} for processing the information. During this processing time, the information used to generate the update ages by $c_i$. When the information provider sends the update to the receiver, the age at the receiver decreases down to the age of the update which is $c_i$, as the communication time between the transmitter and the receiver is negligible.

We model distortion as a monotonically decreasing function of processing time, $c_i$, motivated by the diminishing returns property \cite{Bastopcu18a}. We consider exponentially and inverse linearly decaying distortion functions as examples. In particular, inverse linearly decaying distortion function arises in sensor networking applications, where all sensors observe an underlying random variable distorted by independent Gaussian noise, and the information provider combines sensor observations linearly to minimize the mean squared error (see Section~\ref{sect:sys_model}).

In this paper, we determine age-optimum updating schemes for a system with a distortion constraint on each update. We are given a total time duration over which the average age is calculated $T$, the total number of updates $N$, the maximum allowed distortion as a function of the current age $f(y)$, and the distortion function as a function of the processing time $D(c)$. We solve for the optimum request times for the updates at the receiver and the optimum processing times of the updates at the transmitter, to minimize the overall age.

In this work, we consider the general case where the distortion constraint is a function of the processing time at the transmitter and the current age at the receiver.\footnote{In the conference version of this work in \cite{Bastopcu19}, we considered the simpler case where the distortion constraint was a function of the processing time only, i.e., it was not a function of the current age.} Distortion function is always monotonically decreasing with the processing time. Regarding the dependence of the distortion constraint on the current age at the receiver, we consider three different scenarios: First, as in \cite{Bastopcu19}, distortion constraint is constant (independent of the current age), second, the distortion constraint is inversely proportional with the current age, and third, the distortion constraint is proportional with the current age. The second case is motivated by the following observation: If the age at the receiver is high, the receiver may want to receive a high quality update, i.e., an update with low distortion, to replace its current information with more accurate information. In this case a high age implies a low desired distortion, hence, age and distortion constraints are inversely proportional. The third case is motivated by the following observation: If the age at the receiver is high, the receiver may want to receive a quick update, i.e., an update with high distortion, to replace its current information with a fresh information. In this case, the receiver trades its obsolete but high quality update with a fresh but low quality update. This may be desirable in applications where the freshness of information matters more than the quality of the information. Therefore, in this work, we consider the cases where the distortion constraint is 1) a constant, 2) a decreasing, and 3) an increasing function of the current age.

In this paper, we provide the age-optimal policies by finding the optimum processing times and the optimum update request times. We show that the optimum processing time is always equal to the minimum required processing time that meets the distortion constraint. If there is no active constraint on distortion, i.e., the distortion constraint is high enough, the optimum processing time is equal to zero. We observe three different optimum policies for update request times depending on the level of distortion constraint. When the distortion constraint is large enough except in the case where the distortion function is inversely proportional to the current age, we show that the optimal policy is to request updates with equal inter-update times. When the distortion constraint is relatively large, i.e., the required processing time is relatively small compared to the total time period, it is optimal to request updates regularly following a waiting (request) time after receiving each update, with a longer request time for the first update than others. When the distortion constraint is relatively small, i.e., the required processing time is relatively large compared to the total time period, the optimal policy is to request an update once the previous update is received, i.e., back-to-back, except for a potentially non-zero requesting time for the first update.

\subsection{Related Work}
 
References that are most closely related to our work are \cite{Liu19, Rajaraman18, Alabbasi20, Tang19} which consider the trade-off between service performance and information freshness. \cite{Alabbasi20} emphasizes the difference between service completion time and the age. \cite{Rajaraman18} considers the joint optimization of information freshness, quality of information, and total energy consumption which assumes that the distortion (utility) function follows law of diminishing returns and models the age and energy cost as convex functions. The main contribution of \cite{Rajaraman18} is deriving an online algorithm which is 2-competitive. In our paper, there is no explicit energy constraint, but the total number of updates $N$ for a given total time duration $T$ is limited. Even though we consider the age and quality of the updates, the problem settings are different where we minimize the average age of information, which is inherently non-convex, subject to a distortion constraint for each update. Furthermore, we consider age-dependent distortion constraint which also differentiates our overall work from \cite{Rajaraman18}. 

In \cite{Liu19}, service performance is measured by how \emph{quickly} the provider responds to the queries of the receiver. In \cite{Liu19}, the performance of the system is considered to be the highest when the service provider responds immediately upon a request. In \cite{Liu19}, by responding quickly, the service provider may be using available, but perhaps outdated, information resulting in larger age; on the other hand, if the provider waits for processing new data and responds to the queries a bit later, information of the update may be fresher. Thus, in the model of \cite{Liu19}, processing data degrades quality of service as it worsens response time, but improves the age. In contrast, in our model, processing data improves service performance (the quality of updates), but worsens the age, as the age at the receiver grows while the transmitter processes the data. Thus, the models and trade-offs captured in \cite{Liu19} and here are substantially different.

As we model the distortion as a function of the processing time and the maximum allowed distortion as a function of the instantaneous age, update duration depends on the current age. A similar problem with age-dependent update duration was considered in \cite{Tang19} where the solution for a relaxed and simplified version of the original problem was given. Different from \cite{Tang19}, where only the case in which the update duration is proportional to the current age is considered, here we consider the cases in which the update duration is proportional and inversely proportional with the current age, and we provide exact solutions for both problems. 

Finally, \cite{Ramirez19} and  \cite{Bastopcu20} consider \textit{partial updates} where the information content is smaller compared to full updates, which also resembles trading-off update quality with service time. 

\section{System Model and Problem Formulation} \label{sect:sys_model}

Let $a(t)$ be the instantaneous age at time $t$, with $a(0)=0$. When there is no update, the age increases linearly over time; see Fig.~\ref{Fig:age_eval}. When an update is received, the age at the receiver decreases down to the age of the latest received update. The channel between the information provider and the receiver is assumed to be perfect with zero transmission times, as in e.g., \cite{Feng18a, Feng18c, Bacinoglu15, Bacinoglu17}. However, in order to generate an update, the provider needs to allocate a processing time. For update $i$, the provider allocates $c_i$ amount of processing time. 

We model the distortion function as a monotonically decreasing function of processing time due to the diminishing returns property. For instance, we consider an exponentially decaying distortion function, $D_e$, 
\begin{align}
D_e(c_i) = a\left(e^{-bc_i}-d\right), \label{dist_fnc}
\end{align}
where $d\leq e^{-bc_{max}}$ so that the distortion function is always nonnegative. In addition, we consider an inverse linearly decaying distortion function, $D_\ell$,
\begin{align}
D_\ell(c_i) = \frac{a}{b c_i+d}, \label{dist_fnc2}
\end{align}
which arises in sensor networking applications. In particular, consider a system with $M$ sensors placed in an area, measuring a common random variable $X$ with mean $\mu_X$ and variance $\sigma_X^2$. The measurement at each sensor, $Y_j$, is perturbed by an i.i.d. zero-mean Gaussian noise with variance $\sigma^2$. Information provider uses a linear estimator, $\hat{X} = \sum_{j=1}^{M} w_j Y_j$ to minimize the distortion (mean squared error) defined as $D_\ell= \mathbb{E}[(\hat{X}-X)^2]$. In this model, we assume that the information provider connects to one sensor at a time and spends one unit of time to retrieve the measurement from that sensor. Thus, if the information provider connects to $c_i$ sensors, it spends $c_i$ units of time for processing (i.e., retrieving data) and achieves a distortion of $D_\ell(c_i) = \sigma^2/(c_i+\frac{\sigma^2}{\mu_{X}^2+\sigma_{X}^2})$ for the $i$th update, which has the inverse linearly decaying form in (\ref{dist_fnc2}).

Let $s_i$ be the time interval between the reception time of the $(i-1)$th update and the request time of the $i$th update at the receiver, and let $c_i$ be the processing time of the $i$th update at the transmitter; see Fig.~\ref{Fig:age_eval}. Then, $y_i =s_i+c_{i-1}$ is the time interval between requesting the $(i-1)$th and the $i$th updates; it is also the age at the time of requesting the $i$th update; see Fig.~\ref{Fig:age_eval}. The remaining time after receiving the last update is $s_{N+1}$, i.e., $s_{N+1} = T-\sum_{i=1}^{N}(s_i+c_i)$, and $c_0 =0$. 

We define $f(y_i)$ as the maximum allowed distortion for each update where $y_i$ is the current age. We will start with the case where the maximum allowed distortion is a constant, $f(y_i) =\beta$, i.e., it does not depend on the current age, and then continue with the general case where it explicitly depends on the current age. We consider two sub-cases in the latter case. In the first sub-case, the maximum allowed distortion decreases with the current age, and in the second sub-case, the maximum allowed distortion increases with the current age.      

Our objective is to minimize the average age of information at the information receiver over a total time period $T$, subject to having a desired level of distortion for each update, given that there are $N$ updates. We formulate the problem as,
\begin{align}
\label{problem1}
\min_{\{s_{i}, c_{i} \}}  \quad & \frac{1}{T} \int_{0}^{T} a(t) dt \nonumber \\
\mbox{s.t.} \quad & \sum_{i=1}^{N+1} s_i+c_{i-1} = T \nonumber \\
\quad &D(c_i)\leq f(y_i), \quad i = 1,\dots,N \nonumber \\
\quad & s_i\geq 0, \quad c_i\geq 0,
\end{align}
where $a(t)$ is the instantaneous age, $D(c_i)$ is the distortion function which is monotonically decreasing in $c_i$, and $f(y_i)$ is the maximum allowed distortion function for update $i$ as a function of the current age $y_i$. The distortion function $D(c_i)$ may be $D_e(c_i)$ or $D_\ell(c_i)$ defined above, or any other appropriate distortion function depending on the application. The maximum allowed distortion $f(y_i)$ may be constant, i.e., $f(y_i) =\beta$, or it may be a function of the current age $y_i$. We consider two specific cases where $f(y_i)$ is a decreasing function of $y_i$ and where $f(y_i)$ is an increasing function of $y_i$. Let $A_T \triangleq \int_{0}^{T} a(t) dt$ be the total age. Note that minimizing $\frac{A_T}{T}$ is equivalent to minimizing $A_T$ since $T$ is a known constant.

With these definitions, and using the age evolution curve in Fig.~\ref{Fig:age_eval}, the total age $A_T$ is,  
\begin{align}
A_T = \frac{1}{2}\sum_{i=1}^{N+1}(s_i+c_{i-1})^2+\sum_{i=1}^{N}c_i(s_i+c_{i-1}). \label{Eqn:Age_exp}
\end{align}

In the following section, we provide the optimal solution for the problem defined in (\ref{problem1}) when the maximum allowed distortion is constant.
  
\section{Constant Allowable Distortion}\label{sub_sect:constant}

In this section, we consider the case $f(y_i) = \beta$. Since $D(c_i)$ is a monotonically decreasing function of $c_i$, $D(c_i)\leq \beta$ is equivalent to $c_i\geq c$ where $c=D^{-1}(\beta)$ is a constant. Thus, we replace the distortion constraint given in (\ref{problem1}) with $c_i\geq c$. In addition, we substitute $y_i = s_i+c_{i-1}$ for $i=1,\dots,N+1$. Then, using (\ref{Eqn:Age_exp}), we rewrite the problem in (\ref{problem1}) as,
\begin{align}
\label{problem2}
\min_{\{y_{i}, c_{i} \}}  \quad & \frac{1}{2}\sum_{i=1}^{N+1}y_i^2+\sum_{i=1}^{N}c_iy_i \nonumber \\
\mbox{s.t.} \quad & \sum_{i=1}^{N+1} y_i = T \nonumber \\
\quad &y_1\geq 0, \quad y_i\geq c_{i-1},\quad i=2,\dots,N+1\nonumber \\
\quad &c_i\geq c,\quad i=1,\dots,N.
\end{align}

The optimization problem in (\ref{problem2}) is not convex due to the multiplicative terms involving $c_i$ and $y_i$. We note that $c_i =c$ for $i= 1,\dots,N$ is an optimum selection, since this selection minimizes the second term in the objective function and at the same time yields the largest feasible set for the remaining set of variables (i.e., $y_i$s) in the problem in (\ref{problem2}). Thus, the optimization problem in (\ref{problem2}) becomes,
\begin{align}
\label{problem3}
\min_{\{y_{i} \}}  \quad & \frac{1}{2}\sum_{i=1}^{N+1}y_i^2+\sum_{i=1}^{N}c y_i  \nonumber \\
\mbox{s.t.} \quad & \sum_{i=1}^{N+1} y_i = T \nonumber \\
\quad &y_1\geq 0,\quad y_i\geq c,\quad i=2,\dots,N+1,
\end{align}
which is now only in terms of $y_i$. 

When $\beta=\infty$, and thus, $c=0$ in (\ref{problem3}), i.e., there is no active distortion constraint, the optimal solution is to choose $y_i =\frac{T}{N+1}$ for all $i$. Therefore, for the rest of this section, we consider the case where $\beta<\infty$, and thus, $c>0$.

We write the Lagrangian for the problem in (\ref{problem3}) as,
\begin{align}
\mathcal{L} =& \frac{1}{2}\sum_{i=1}^{N+1}y_i^2+\sum_{i=1}^{N}c y_i-\lambda\left(\sum_{i=1}^{N+1} y_i -T \right)-\sum_{i=2}^{N+1}\theta_i(y_i-c)\nonumber\\
&-\theta_1y_1,
\end{align}
where $\theta_i\geq 0$ and $\lambda$ can be anything. The problem in (\ref{problem3}) is convex. Thus, the KKT conditions are necessary and sufficient for the optimal solution. The KKT conditions are,
\begin{align}
\frac{\partial \mathcal{L}}{\partial y_i} & = y_i+c-\lambda-\theta_i = 0,\quad i=1,\dots,N,\label{KKT2} \\
\frac{\partial \mathcal{L}}{\partial y_{N+1}} & = y_{N+1}-\lambda-\theta_{N+1}= 0. \label{KKT3}
\end{align}
The complementary slackness conditions are,
\begin{align}
\lambda\left(\sum_{i=1}^{N+1} y_i - T \right) & =0,\label{cs1} \\
\theta_1y_1 & = 0 ,\label{cs2}\\
\theta_i(y_i-c) & =0, \quad i=2,\dots,N+1. \label{cs3}
\end{align}

When $y_i>c$ for all $i$, we have $\theta_i =0$ due to (\ref{cs2}) and (\ref{cs3}). Then, from (\ref{KKT2}) and (\ref{KKT3}), we obtain $y_i = \lambda-c$ for $i=1,\dots,N$, and $ y_{N+1} = \lambda$. Since $ \sum_{i=1}^{N+1} y_i =T$, we find $\lambda=\frac{T+Nc}{N+1}$. Thus, the optimal solution becomes,
\begin{align}
y_i &= \frac{T-c}{N+1}, \quad  i=1,\dots,N,\\
y_{N+1}& = \frac{T+Nc}{N+1}.
\end{align}
In order to have $y_i>c$, we need $T>(N+2)c$. Viewing this condition from the perspective of $c$, this is the case when $c$ is small in comparison to $T$. Therefore, we note that, in this case, when minimum processing time, $c$, is relatively small, the optimal policy is to choose $y_i$ as equal as possible except for $y_{N+1}$. When $c$ becomes larger compared to $T$, $y_i-c$ decreases. Specifically, when $T = (N+2)c$, $y_i=c$ for $i = 1,\dots,N$.

In the remaining case, i.e., when $T < (N+2)c$, $y_1<c$ and $y_{N+1}>c$, we have $\theta_1 =0$ and $\theta_{N+1}=0$ by (\ref{cs2}) and (\ref{cs3}). Then, by solving $y_i = \lambda-c$, $y_{N+1} = \lambda$, and $ \sum_{i=1}^{N+1} y_i = T$, we obtain,
\begin{align}
y_1 &= \frac{T-Nc}{2}, \\
y_i &= c,\quad i = 2,\dots,N,\\
y_{N+1} &= \frac{T-(N-2)c}{2}.
\end{align}
Since $y_1>0$, we need $Nc<T$. Thus, this solution applies when $Nc<T\leq (N+2)c$. 

Finally, when $T=Nc$, the optimal solution becomes,
\begin{align}
y_1 &= 0,\\
y_i &= c,\quad i = 2,\dots,N+1.
\end{align}

\begin{figure*}[t]
	\begin{center}
		\subfloat[]{%
			\includegraphics[width=0.40\linewidth]{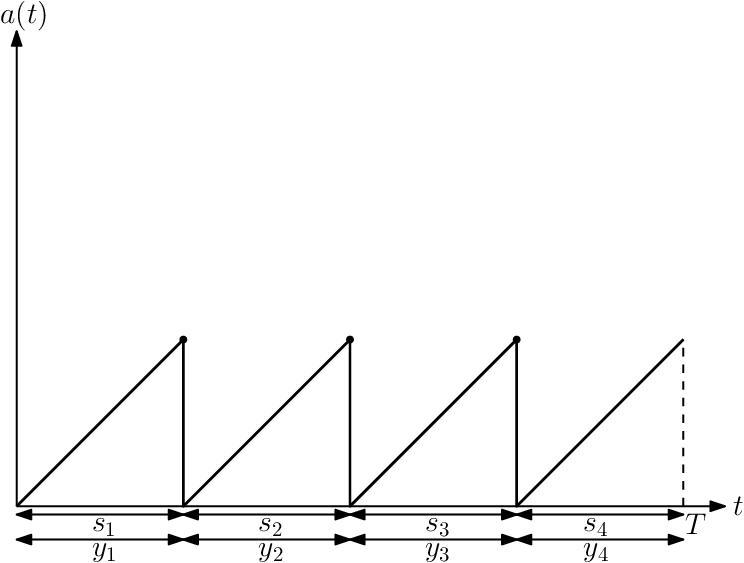}}\hfil
		\subfloat[]{%
			\includegraphics[width=0.40\linewidth]{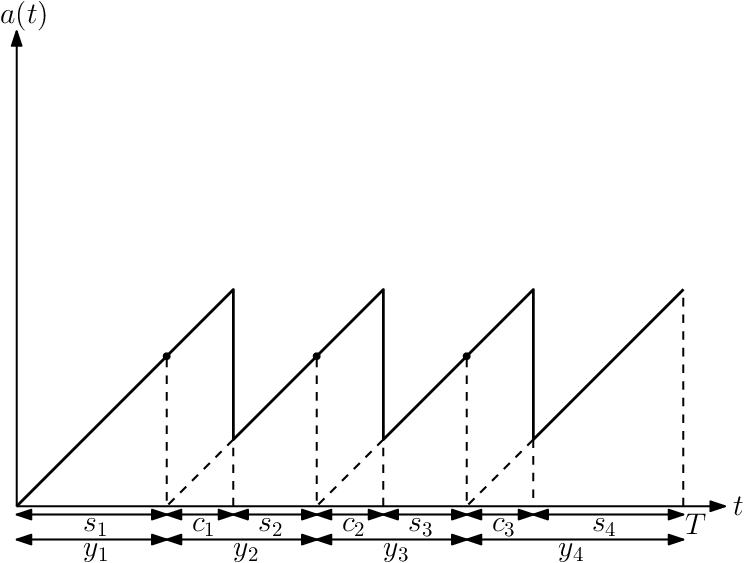}} \\
		\subfloat[]{%
			\includegraphics[width=0.40\linewidth]{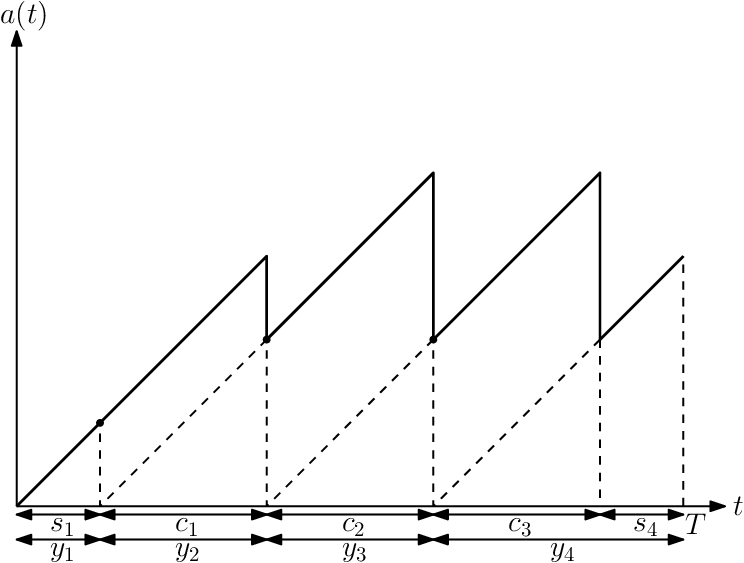}}\hfil
		\subfloat[]{%
			\includegraphics[width=0.40\linewidth]{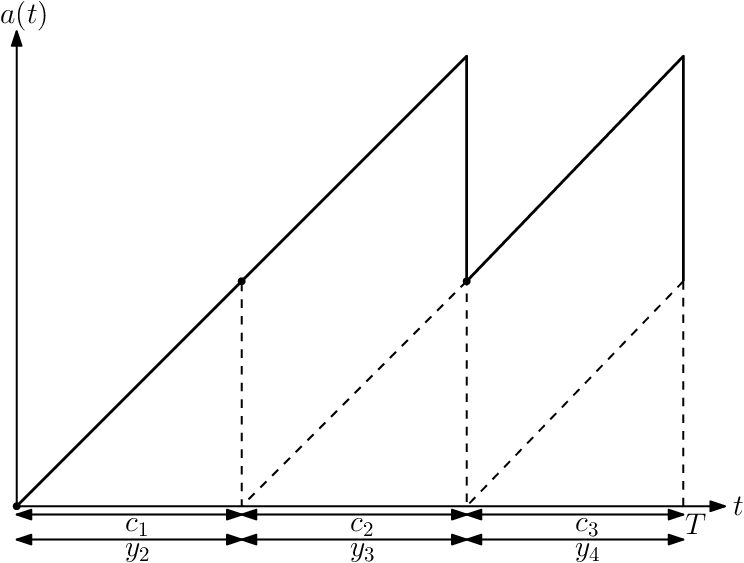}}
		\caption{Evolution of $a(t)$ with optimal update policies when the distortion function does not depend on the current age in the case of (a) $c=0$, (b) $c>0$ and $(N+2)c<T$, (c) $Nc<T\leq (N+2)c$, (d) $T=Nc$.}
		\label{Fig:Age_eval_constant_age}
	\end{center}
	\vspace{-4mm}
\end{figure*}

In summary, when $c=0$, i.e., we do not have any distortion constraints, then the optimal solution is to update in every $\frac{T}{N+1}$ units of time, i.e., $y_i =\frac{T}{N+1}$ for all $i$. When $c>0$ but, relatively small compared to $T$, i.e., $(N+2)c<T$, the optimal solution is to wait for $\frac{T-c}{N+1}$ to request the first update. For the remaining updates, the receiver waits for $\frac{T-(N+2)c}{N+1}$ time to request another update after the previous update is received. After requesting $N$ updates, the optimal policy is to let the age grow for the remaining $\frac{T-c}{N+1}$ units of time. When $c$ becomes large compared to $T$, i.e., $Nc<T\leq (N+2)c$, the optimal policy is to wait for $\frac{T-Nc}{2}$ to request the first update and request the remaining updates as soon as the previous update is received, i.e., back-to-back. After updating $N$ times, we let the age grow for the remaining $\frac{T-Nc}{2}$ units of time. Finally, when $T=Nc$, the optimal policy is to request the first update at $t=0$ and request the remaining updates as soon as the previous update is received, i.e., back-to-back. We note that when $Nc>T$, there is no feasible policy. The possible optimal policies are shown in Fig.~\ref{Fig:Age_eval_constant_age}. 

In the following section, we provide the optimal solution for the problem defined in (\ref{problem1}) when the maximum allowed distortion is age-dependent. 

\section{Age-Dependent Allowable Distortion}\label{sub_sect:age}

In this section, we consider the case where the maximum allowed distortion $f(y_i)$ depends explicitly on the instantaneous age $y_i$. As motivated in the introduction section, this dependence may take different forms. In particular, depending on the application, $f(y_i)$ may be a decreasing or an increasing function of $y_i$. In the following two sub-sections, we consider two sub-cases: when $f(y_i)$ is inversely proportional to $y_i$ and when $f(y_i)$ is proportional to $y_i$.  

\subsection{Allowable Distortion is Inversely Proportional to the Instantaneous Age}\label{Subsect:prop}

We consider the case where $f(y_i)$ is a decreasing function of $y_i$. Since the distortion function $D(c_i)$ is a decreasing function of the processing time $c_i$, the distortion constraint for each update, i.e., $D(c_i)\leq f(y_i)$, becomes $c_i\geq D^{-1}(f(y_i))$ where $D^{-1}(\cdot)$ is the inverse function of the distortion function. As $f(y_i)$ is a decreasing function of $y_i$, the minimum required processing time $D^{-1}(f(y_i))$ is an increasing function of the current age $y_i$, i.e., we have $D^{-1}(f(y_j))\geq D^{-1}(f(y_i))$ for all $y_j\geq y_i$. In general, $D^{-1}(f(y_i))$ function can be arbitrary depending on the selections of $D(c_i)$ and $f(y_i)$. However, in order to make the analysis tractable, in this paper, we focus on a particular case where the distortion constraint for each update in (\ref{problem1}), i.e., $D(c_i)\leq f(y_i)$, implies $c_i\geq \alpha y_i$, where $\alpha$ is a positive constant. An example for this case is obtained, if we consider the inverse linearly decaying distortion function, $D_\ell(c_i) = \frac{a}{b c_i+d}$ in (\ref{dist_fnc2}), and use an inverse linearly decaying allowable distortion function $f(y_i) = \frac{a}{\kappa y_i+d}$. 

The optimization problem in (\ref{problem1}) in this case becomes,
\begin{align}
\label{problem4}
\min_{\{y_{i}, c_{i} \}}  \quad & \frac{1}{2}\sum_{i=1}^{N+1}y_i^2+\sum_{i=1}^{N}c_iy_i \nonumber \\
\mbox{s.t.} \quad & \sum_{i=1}^{N+1} y_i = T \nonumber \\
\quad &y_1\geq 0, \quad y_i\geq c_{i-1},\quad i=2,\dots,N+1\nonumber \\
\quad &c_i\geq 0, \quad c_i\geq \alpha y_i,\quad i=1,\dots,N.
\end{align}

In the following lemma, we show that the processing time for each update should be equal to the minimum required time to satisfy the distortion constraint, i.e.,  $c_i = \alpha y_i$, for all $i$. 

\begin{lemma}\label{Lemma:age_dec_1}
	In the age-optimal policy, processing time for each update is equal to the minimum required time which meets the distortion constraint with equality, i.e., $c_i = \alpha y_i$ for all $i$.
\end{lemma}

\begin{Proof}
	Let us assume for contradiction that there exists an optimal policy such that $c_j>\alpha y_j$ for some $j$. Then, we find another feasible policy denoted by $\{s_i',c_i'\}$ such that $c_j' = c_j-\epsilon$, $s_{j+1}' = s_{j+1}+\epsilon$ and $y_{j+1}' = s_{j+1}'+c_j' = y_{j+1}$. Since $c_j>\alpha y_j$, we can always choose sufficiently small $\epsilon$ so that we have $c_j'\geq \alpha y_j'$ for the new policy. We have $y_i = y_i'$ for all $i$ and $c_i =c_i'$ for $i\neq j$ which means that in the new policy, we keep all other variables the same except for $c_j'$ and $s_{j+1}'$. Inspecting the objective function of (\ref{problem4}), we note that in the new policy, the age is decreased by $\epsilon y_j$. Since the new policy with $\{s_i',c_i'\}$ achieves a smaller age, we reach a contradiction. Therefore, in the age-optimal policy, we must have $c_i = \alpha y_i$, for all $i$.   
\end{Proof}

We remark that Lemma~\ref{Lemma:age_dec_1} provides an alternative proof for the fact that $c_i$ must be such that $c_i =c$ in (\ref{problem2}). We argued this briefly after (\ref{problem2}) based on the observation that this selection minimizes the objective function and enlarges the feasible set. 

\begin{figure*}[t]
	\begin{center}
		\subfloat[]{%
			\includegraphics[width=0.4\linewidth]{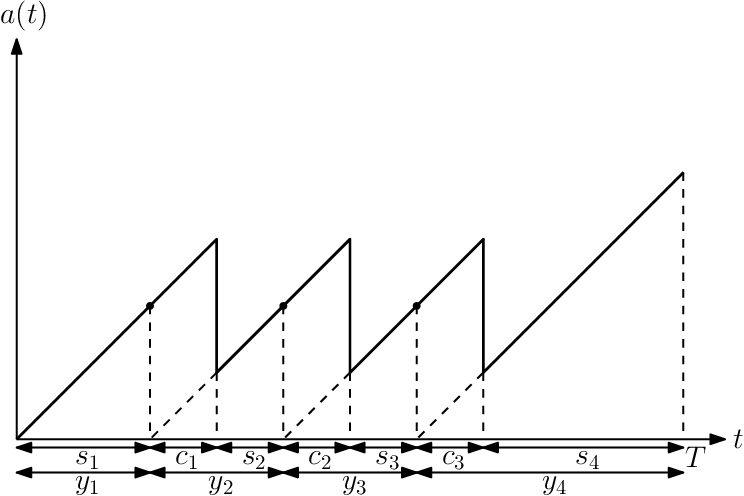}} \hfil
		\subfloat[]{%
			\includegraphics[width=0.40\linewidth]{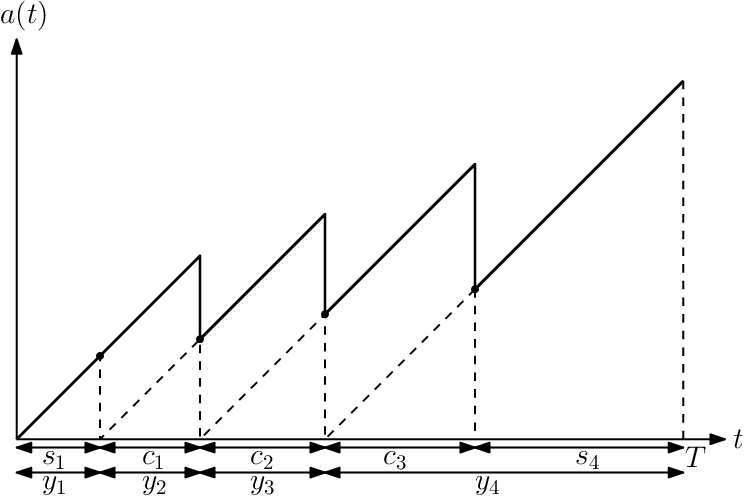}}
		\caption{Age evolution at the receiver when $f(y_i)$ is inversely proportional to the current age for (a) $\alpha\leq1 $ and (b) $\alpha >1$.}
		\label{Fig:age_eval_dld_age_dec}
	\end{center}
	\vspace{-4mm}
\end{figure*}

Using Lemma~\ref{Lemma:age_dec_1}, we let $c_i = \alpha y_i$, and rewrite (\ref{problem4}) as,  
\begin{align}
\label{problem4_mod}
\min_{\{y_{i} \}}  \quad & \left(\frac{1}{2}+\alpha \right)\sum_{i=1}^{N}y_i^2+\frac{1}{2}y_{N+1}^2 \nonumber \\
\mbox{s.t.} \quad & \sum_{i=1}^{N+1} y_i = T \nonumber \\
\quad &y_1\geq 0, \quad y_i\geq \alpha y_{i-1},\quad i=2,\dots,N+1,
\end{align}
which is only in terms of $y_i$. 

We write the Lagrangian for the problem in (\ref{problem4_mod}) as,
\begin{align}
\mathcal{L} =& \left(\frac{1}{2}+\alpha \right)\sum_{i=1}^{N}y_i^2+\frac{1}{2}y_{N+1}^2-\lambda\left(\sum_{i=1}^{N+1} y_i -T \right)-\beta_1y_1\nonumber\\
&-\sum_{i=2}^{N+1}\beta_i(y_i-\alpha y_{i-1}),
\end{align}
where $\beta_i\geq 0$ and $\lambda$ can be anything. The problem in (\ref{problem4_mod}) is convex. Thus, the KKT conditions are necessary and sufficient for the optimal solution. The KKT conditions are, 
\begin{align}
\frac{\partial \mathcal{L}}{\partial y_i}  =& (1+2\alpha)y_i-\lambda-\beta_i+\alpha \beta_{i+1} = 0,\quad i=1,\dots,N,\label{KKT1_1} \\
\frac{\partial \mathcal{L}}{\partial y_{N+1}} =& y_{N+1}-\lambda-\beta_{N+1} = 0. \label{KKT2_2}
\end{align}
The complementary slackness conditions are,
\begin{align}
\lambda\left(\sum_{i=1}^{N+1} y_i - T \right) & =0,\label{cs1_2} \\
\beta_1y_1 & = 0, \label{cs2_2}\\
\beta_i(y_i-\alpha y_{i-1}) & =0, \quad i=2,\dots,N+1. \label{cs3_2}
\end{align}

First, we consider the case where $s_i>0$ and $c_i>0$ for all $i$. Then, we have $y_1>0$ and $y_i>\alpha y_{i-1}$ for all $i= 2,\dots, N+1$. The former statement follows because $y_1=s_1>0$, and the latter statement follows because $y_i= \alpha c_i$ due to Lemma~\ref{Lemma:age_dec_1} and $y_i = s_i+c_{i-1} = s_i+\alpha y_{i-1}>\alpha y_{i-1}$ since $s_i>0$. Thus, from (\ref{cs2_2})-(\ref{cs3_2}), we have $\beta_i =0$ for all $i$. By using (\ref{KKT1_1})-(\ref{KKT2_2}), we have $y_i = \frac{\lambda}{2\alpha+1}$ for $i=1,\dots,N$, and $ y_{N+1} = \lambda$. Since $ \sum_{i=1}^{N+1} y_i = T $ from (\ref{cs1_2}), we find $\lambda = \frac{(2\alpha+1)T}{N+2\alpha+1}$. Thus, the optimal solution in this case is, 
\begin{align}
y_i &= \frac{T}{N+2\alpha+1},\quad i = 1,\dots,N,\label{opt_yi}\\
y_{N+1} &= \frac{(2\alpha+1)T}{N+2\alpha+1}.\label{opt_y_N+1}
\end{align}
In order to satisfy $y_i> \alpha y_{i-1}$, we need $\alpha<1$. A typical age evolution curve for $\alpha<1$ is shown in Fig.~\ref{Fig:age_eval_dld_age_dec}(a). When $\alpha=1$, we note that the optimal solution follows (\ref{opt_yi}) and (\ref{opt_y_N+1}), but $y_i = \alpha y_{i-1}$ for $i =2,\dots, N$.

Next, we find the optimal solution for $\alpha > 1$. If we have only the total time constraint, then the optimal solution is to choose $y_i$s equal for $i=1,\dots,N$. Since $\alpha>1$, we cannot choose $y_i$s equal due to $y_i\geq \alpha y_{i-1}$ constraints. In the following lemma, we prove that  when $\alpha>1$, $y_i =\alpha y_{i-1}$ for $i = 2,\dots,N$.
 
\begin{lemma}\label{Lemma:age_dec_2}
	When $\alpha>1$, we have $y_i =\alpha y_{i-1}$ for $i = 2,\dots, N$. 
\end{lemma} 

\begin{Proof}
	Assume for contradiction that there exists an age-optimal policy with $y_j>\alpha y_{j-1}$ for some $j\in \{2,\dots,N\}$. From (\ref{cs3_2}), we have $\beta_j = 0$. From (\ref{KKT1_1}), we get $y_j = \frac{\lambda-\alpha B_{j+1}}{2\alpha +1}$ and $y_{j-1}= \frac{\lambda+\beta_{j-1}}{2\alpha+1}$. Since $y_j\geq 0$, we must have $\lambda\geq 0$. By using $y_j > \alpha y_{j-1}$, we must have $(1-\alpha)\lambda > \alpha (\beta_{j+1}+\beta_{j-1})$. Since $\alpha>1$ and $\lambda\geq 0$, this implies $ (1-\alpha)\lambda\leq 0$, which further implies $ \alpha (\beta_{j+1}+\beta_{j-1})<0 $. However, this inequality cannot be satisfied since $\beta_i\geq0$ for all $i$. Thus, we reach a contradiction and in the age-optimal policy, we must have $y_i = \alpha y_{i-1}$ for $i = 2,\dots, N$.          
\end{Proof}

Then, the optimal policy is in the form of $y_i = \alpha^{i-1} \eta$ for $i = 1,\dots, N$ and $y_{N+1} = T-\sum_{i=1}^{N}y_i$ where $\eta$ is a constant. We write the total age in terms of $\eta$ as, 
\begin{align}\label{age_dec_opt_alp>1}
A_T(\eta) = &\left( \frac{1}{2}+\alpha\right)\eta^2\left( \frac{\alpha^{2N}-1}{\alpha^2-1}\right)\nonumber\\
&+\frac{1}{2}\left( T-\left(\frac{\alpha^N-1}{\alpha-1}\right)\eta\right)^2. 
\end{align}
In order to find the optimal $\eta$, we differentiate (\ref{age_dec_opt_alp>1}), which is quadratic in $\eta$, with respect to $\eta$ and equate to zero. We find the optimal solution for $\alpha>1$
as, 
\begin{align}
y_1 =& \frac{T(\alpha^{N+2}-\alpha^N-\alpha^2+1)}{2(\alpha^{2N+2}-\alpha^{N+1}-\alpha^N-\alpha^2+\alpha+1)},\label{age_opt_y1}\\
y_i =& \alpha^{i-1}y_{i-1}, \quad i=2,\dots,N, \\
y_{N+1} =& T-\sum_{i=1}^{N}y_i. \label{age_opt_y_son}
\end{align}
A typical age evolution curve for $\alpha>1$ is shown in Fig.~\ref{Fig:age_eval_dld_age_dec}(b).  

\subsection{Allowable Distortion is Proportional to the Instantaneous Age}

We consider the case where $f(y_i)$ is an increasing function of $y_i$. Similar to Section~\ref{Subsect:prop}, the distortion constraint for each update, i.e., $D(c_i)\leq f(y_i)$, is equivalent to $c_i\geq D^{-1}(f(y_i))$. As $f(y_i)$ is an increasing function of $y_i$, the minimum required processing time $D^{-1}(f(y_i))$ is a decreasing function of the current age $y_i$, i.e., we have $D^{-1}(f(y_j))\leq D^{-1}(f(y_i))$ for all $y_j\geq y_i$. Even though $D^{-1}(f(y_j))$ can be arbitrary, in this paper, in order to make the analysis tractable, we focus on a specific case where the distortion constraint for each update in (\ref{problem1}), i.e., $D(c_i)\leq f(y_i)$, implies $c_i\geq c-\alpha y_i$. In this section, we assume $\alpha <\frac{1}{2}$. An example of this case is obtained, if we consider the inverse linearly decaying distortion, $D_\ell(c_i) = \frac{a}{b c_i+d}$ in (\ref{dist_fnc2}), and use $f(y_i) = \frac{a}{u-\kappa y_i}$. Thus, while the distortion constraint in Section~\ref{Subsect:prop} was $c_i\geq \alpha y_i$, the distortion constraint in this section is $c_i\geq c-\alpha y_i$. 

The optimization problem in (\ref{problem1}) in this case becomes,
\begin{align}
\label{problem6}
\min_{\{y_{i}, c_{i} \}}  \quad & \frac{1}{2}\sum_{i=1}^{N+1}y_i^2+\sum_{i=1}^{N}c_iy_i \nonumber \\
\mbox{s.t.} \quad & \sum_{i=1}^{N+1} y_i = T \nonumber \\
\quad &y_1\geq 0, \quad y_i\geq c_{i-1},\quad i=2,\dots,N+1\nonumber \\
\quad &c_i\geq 0, \quad c_i\geq c-\alpha y_i,\quad i=1,\dots,N.
\end{align}  

In the following lemma, we show that the processing time for each update should be equal to the minimum processing time which satisfies the distortion constraint, i.e., $c_i = (c-\alpha y_{i})^+$ for $i=1,\dots,N$, where $(x)^+ = \max\{0,x\}$.

\begin{lemma}\label{Lemma:age_inc_1}
	In the age-optimal policy, processing time for each update is equal to the minimum required time which meets the distortion constraint with equality, i.e., $c_i = (c-\alpha y_i)^+$, for all $i$.
\end{lemma}  

\begin{Proof}
	Let us assume for contradiction that there exists an optimal policy such that $c_j>c-\alpha y_j$ for some $j$. If $y_j<\frac{c}{\alpha}$, then we find another feasible policy denoted by $\{s_i',c_i'\}$ such that $c_j' = c_j-\epsilon$, $s_{j+1}' = s_{j+1}+\epsilon$ and $y_{j+1}' = s_{j+1}'+c_j' = y_{j+1}$. Since $c_j>c-\alpha y_j$, we can always choose sufficiently small $\epsilon$ so that we have $c_j'\geq c-\alpha y_j'$ for the new policy. We have $y_i = y_i'$ for all $i$ and $c_i =c_i'$ for $i\neq j$ which means that in the new policy, we keep all other variables the same except $c_j'$ and $s_{j+1}'$. We note that in the new policy, age is decreased by $\epsilon y_j$. Since the new policy with $\{s_i',c_i'\}$ achieves a smaller age, we reach a contradiction. Therefore, in the age-optimal policy, we must have $c_i = c- \alpha y_i$ for all $i$ when $y_i<\frac{c}{\alpha}$. If $y_j\geq \frac{c}{\alpha}$, then $c_j\geq0$ is the only constraint on $c_j$. If $c_j>0$, we can similarly argue that decreasing $c_j$ further reduces the age until $c_j$ becomes zero. Thus, we reach a contradiction and when $y_j \geq \frac{c}{\alpha}$, in the optimal solution, we must have $c_j = 0$. By combining these two parts, we conclude that in the optimal policy, we must have $c_i = (c-\alpha y_i)^+$, for $i=1,\dots,N$.       
\end{Proof}

\begin{figure*}[t]
	\begin{center}
		\subfloat[]{%
			\includegraphics[width=0.40\linewidth]{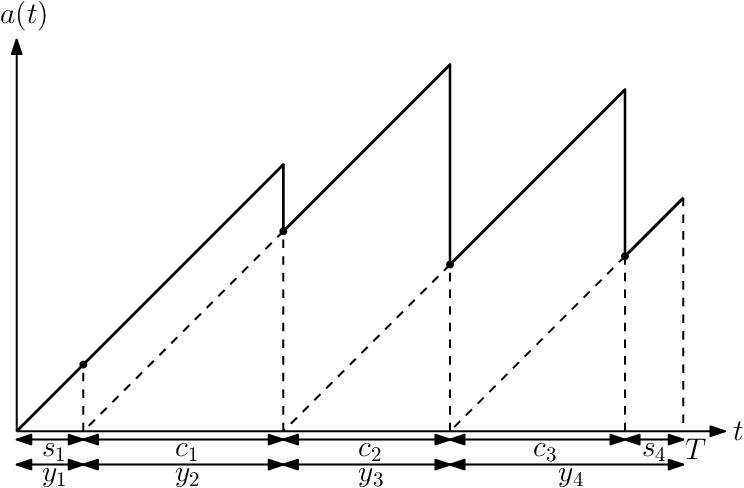}}\hfil
		\subfloat[]{%
			\includegraphics[width=0.40\linewidth]{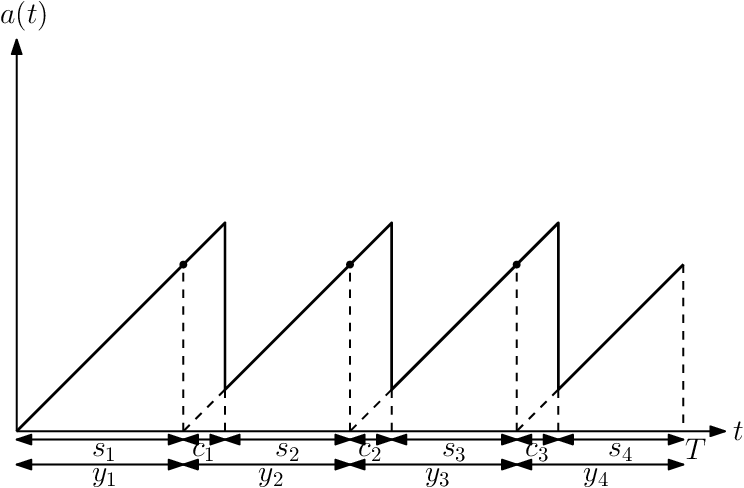}} \\
		\subfloat[]{%
			\includegraphics[width=0.40\linewidth]{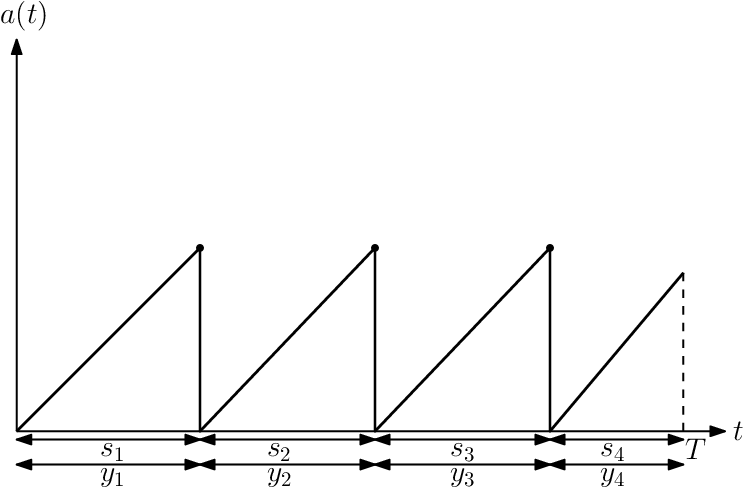}}\hfil
		\subfloat[]{%
			\includegraphics[width=0.40\linewidth]{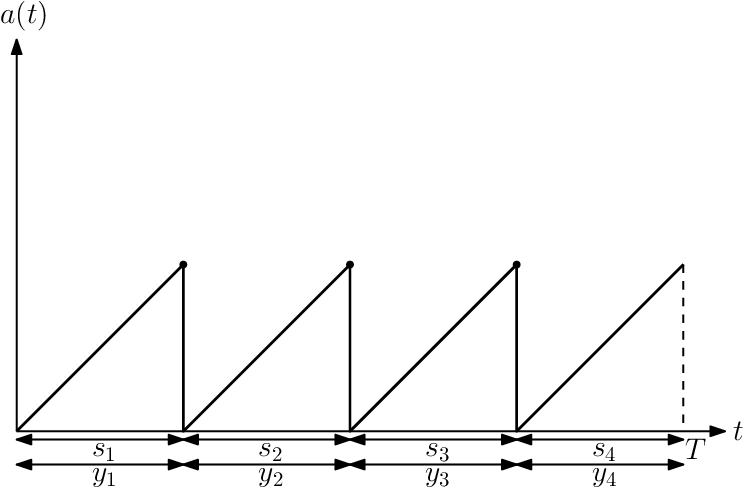}}
		\caption{Age evolution at the receiver when the distortion function is proportional to the current age for (a) $T\leq \left(\frac{N+2-\alpha}{1+\alpha}\right)c$, (b) $\left(\frac{N+2-\alpha}{1+\alpha}\right)c<T<\left(\frac{N+1-\alpha}{\alpha}\right)c$, (c) $ \frac{(N+1-\alpha)c}{\alpha}\leq T<\frac{(N+1)c}{\alpha}$, (d) $\frac{(N+1)c}{\alpha}\leq T$.}
		\label{Fig:age_eval_dld_age_inc}
	\end{center}
	\vspace{-4mm}
\end{figure*}

Using Lemma~\ref{Lemma:age_inc_1}, we let $c_i = (c-\alpha y_i)^+$, and rewrite (\ref{problem6}), 
\begin{align}
\label{problem5}
\min_{\{y_{i} \}}  \quad & \frac{1}{2}\sum_{i=1}^{N+1}y_i^2+\sum_{i=1}^{N}y_i(c-\alpha y_i)^+ \nonumber \\
\mbox{s.t.} \quad & \sum_{i=1}^{N+1} y_i = T \nonumber \\
\quad &y_1\geq 0, \quad y_i\geq (c-\alpha y_{i-1})^+,\quad i=2,\dots,N+1,
\end{align}
which is only in terms of $y_i$. 

Next, we provide the optimal solution for the case where $y_i<\frac{c}{\alpha}$ for $i= 1,\dots, N$. The problem in (\ref{problem5}) becomes, 
\begin{align}
\label{problem5_mod}
\min_{\{y_{i} \}}  \quad & \left(\frac{1}{2}-\alpha \right)\sum_{i=1}^{N}y_i^2+\sum_{i=1}^{N}cy_i+\frac{1}{2}y_{N+1}^2 \nonumber \\
\mbox{s.t.} \quad & \sum_{i=1}^{N+1} y_i = T \nonumber \\
\quad &y_1\geq 0, \quad y_i\geq c-\alpha y_{i-1},\quad i=2,\dots,N+1.
\end{align}
We write the Lagrangian for the problem in (\ref{problem5_mod}) as,
\begin{align}
\mathcal{L} =& \left(\frac{1}{2}-\alpha \right)\sum_{i=1}^{N}y_i^2+\sum_{i=1}^{N}cy_i+\frac{1}{2}y_{N+1}^2-\lambda\left(\sum_{i=1}^{N+1} y_i -T \right)\nonumber\\
&-\beta_1y_1-\sum_{i=2}^{N+1}\beta_i(y_i+\alpha y_{i-1}-c),
\end{align}
where $\beta_i\geq 0$ and $\lambda$ can be anything. The problem in (\ref{problem5_mod}) is convex since $\alpha <\frac{1}{2}$. Thus, the KKT conditions are necessary and sufficient for the optimal solution. The KKT conditions are, 
\begin{align}
\frac{\partial \mathcal{L}}{\partial y_i}  =& (1-2\alpha)y_i+c-\lambda-\beta_i-\alpha \beta_{i+1} = 0,\nonumber\\& i=1,\dots,N,\label{KKT2_1} \\
\frac{\partial \mathcal{L}}{\partial y_{N+1}}  =& y_{N+1}-\lambda-\beta_{N+1} = 0. \label{KKT2_3}
\end{align}
The complementary slackness conditions are,
\begin{align}
\lambda\left(\sum_{i=1}^{N+1} y_i - T \right) & =0,\label{cs1_3} \\
\beta_1y_1 & = 0, \label{cs2_3}\\
\beta_i(y_i+\alpha y_{i-1}-c) & =0, \quad i=2,\dots,N+1. \label{cs3_3}
\end{align}
When $y_1>0$ and $y_i>c-\alpha y_{i-1}$, for $i= 2,\dots, N+1$, from (\ref{cs2_3}) and (\ref{cs3_3}), we have $\beta_i = 0$ for all $i$. Then, by using (\ref{KKT2_1}) and (\ref{KKT2_3}), we have $y_i = \frac{\lambda-c}{1-2\alpha}$, for $i= 1,\dots, N$ and $y_{N+1} = \lambda$. From (\ref{cs1_3}), we find $\lambda = \frac{(1-2\alpha)T+Nc}{N+1-2\alpha}$ which gives, 
\begin{align}
 y_i =& \frac{T-c}{N+1-2\alpha},\quad i=1,\dots, N,\label{y_1_opt_case_2} \\ 
 y_{N+1} =& \frac{(1-2\alpha)T+Nc}{N+1-2\alpha}.\label{y_N+1_opt_case_2}
\end{align}
A typical age evolution curve is shown in Fig.~\ref{Fig:age_eval_dld_age_inc}(b). In order to satisfy $y_1>0$, $y_i> c-\alpha y_{i-1}$ for $i = 2,\dots,N+1$ and $y_i<\frac{c}{\alpha}$ for $i = 1,\dots, N$, we need $\left(\frac{N+2-\alpha}{1+\alpha}\right)c<T<\left(\frac{N+1-\alpha}{\alpha}\right)c$. Viewing this conditions in terms of $T$, when $T$ is closer to the lower boundary, i.e., $\left(\frac{N+2-\alpha}{1+\alpha}\right)c<T$, we see that $y_i>c-\alpha y_{i-1}$ for $i=2,\dots,N$ gets tighter. When $T$ is closer to the upper boundary, we see that $y_i<\frac{c}{\alpha}$, for $i=1,\dots,N$ gets tighter.
 
We first identify the optimal solution when $T\leq \left(\frac{N+2-\alpha}{1+\alpha}\right)c $. In the following lemma, we show that when $T\leq \left(\frac{N+2-\alpha}{1+\alpha}\right)c$, we have $y_i = c-\alpha y_{i-1}$, for $i=2,\dots,N$.
 
\begin{lemma}\label{Lemma:age_inc_2}
	In the age-optimal policy, when $T\leq \left(\frac{N+2-\alpha}{1+\alpha}\right)c$, we have $y_i = c-\alpha y_{i-1}$, for $i=2,\dots,N$.  
\end{lemma}  

\begin{Proof}
	 We note that increasing $c$ increases the cost of increasing $y_i$ for $i=1,\dots,N$ in the objective function in (\ref{problem5_mod}). Thus, increasing $c$ yields decreasing optimal values for $y_i$ for $i=1,\dots,N$. We note from (\ref{y_1_opt_case_2}) that
	 \begin{align}
	 \lim\limits_{T\rightarrow \left(\frac{N+2-\alpha}{1+\alpha}\right)c}y_i = \frac{c}{1+\alpha},    
	 \end{align}
	 for $i=1,\dots,N$. Thus, when $\left(\frac{1+\alpha}{N+2-\alpha}\right)T\leq c$, we have $y_i \leq \frac{c}{1+\alpha}$ for $i=1,\dots,N$. Then, we have $y_i+\alpha y_{i-1} \leq c$ for $i=2,\dots,N$. Due to the distortion constraint in the optimization problem in (\ref{problem5_mod}), we also have $y_i+\alpha y_{i-1}\geq c$ for $i=2,\dots,N+1$. Thus, when $ T\leq \left(\frac{N+2-\alpha}{1+\alpha}\right)c$, we must have $y_i = c-\alpha y_{i-1}$, for $i=2,\dots,N$.              	
\end{Proof}

Therefore, we show in Lemma~\ref{Lemma:age_inc_2} that when $T\leq \left(\frac{N+2-\alpha}{1+\alpha}\right)c$, the optimal policy has the following structure,
\begin{align}
y_1 =& \eta,\label{y_1_opt} \\
y_i  =& c\sum_{j=1}^{i-1}(-\alpha)^{j-1}+(-\alpha)^{i-1}\eta, \quad i=2,\dots,N, \\ 
y_{N+1} =& T-\sum_{i=1}^{N}y_i.\label{y_N+1_opt}
\end{align}
In order to find the optimal $\eta$ which minimizes the age, we substitute (\ref{y_1_opt})-(\ref{y_N+1_opt}) in the objective function in (\ref{problem5_mod}), differentiate the age with respect to $\eta$, and equate to zero.

A typical age evolution curve is shown in Fig.~\ref{Fig:age_eval_dld_age_inc}(a). We note that when we increase $c$ sufficiently, $y_1$ becomes zero. At this point, $y_1\geq 0$ and $y_i\geq c-\alpha y_{i-1}$ for $i =2,\dots, N$ are satisfied with equality. If we further increase $c$, the last feasibility constraint, $y_{N+1}\geq c-y_N$, becomes tight and the optimal solution is $y_1 =0$, $y_i = c-\alpha y_{i-1}$ for $i=2,\dots,N+1$. If we increase $c$ further, there is no feasible solution.

Next, we find the optimal solution when $T$ is relatively large, i.e., $\left(\frac{N+1-\alpha}{\alpha} \right)c<T $. With an argument similar to that in Lemma~\ref{Lemma:age_inc_2}, if $c$ becomes smaller compared to $T$, the optimal value of $y_i$ for $i = 1,\dots,N$ increases. We note that when $\lim\limits_{T\rightarrow \left(\frac{N+1-\alpha}{\alpha} \right)c}y_i = \frac{c}{\alpha}$ for $i= 1,\dots,N$. Thus, when $  \left(\frac{N+1-\alpha}{\alpha} \right)c<T$, we have $c-\alpha y_i<0$ for $i=1,\dots,N$. Then, the problem in (\ref{problem5}) becomes, 
\begin{align}
\label{problem5_mod2}
\min_{\{y_{i} \}}  \quad & \frac{1}{2}\sum_{i=1}^{N+1}y_i^2 \nonumber \\
\mbox{s.t.} \quad & \sum_{i=1}^{N+1} y_i = T \nonumber \\
\quad & y_{N+1}\geq 0, \quad y_i\geq \frac{c}{\alpha}, \quad i=1,\dots,N.
\end{align}
We note that the problem in (\ref{problem5_mod2}) is convex. Thus, the KKT conditions are necessary and sufficient for the optimal solution. After writing the KKT conditions, we observe two different optimal solution structures. When $T$ is sufficiently large, we have $y_i>\frac{c}{\alpha}$ for all $i$. Then, the optimal solution is $y_i = \frac{T}{N+1}$ for all $i$. A typical age evolution curve is shown in Fig.~\ref{Fig:age_eval_dld_age_inc}(d). We need $T\geq \frac{(N+1)c}{\alpha}$ for the feasibility of the solution. When $ \frac{(N+1-\alpha)c}{\alpha}\leq T<\frac{(N+1)c}{\alpha}$, we have $y_i =\frac{c}{\alpha}$ for $i=1,\dots,N$ and $y_{N+1} = T-\sum_{i=1}^{N}y_i$. A typical age evolution curve is shown in Fig.~\ref{Fig:age_eval_dld_age_inc}(c).          

\section{Numerical Results} \label{sect:num_res}

In this section, we provide numerical results for the problems solved in Sections~\ref{sub_sect:constant} and \ref{sub_sect:age}. First, in the following subsection, we provide numerical results for the case where the maximum allowed distortion function is a constant.    

\begin{figure*}[t]
	\begin{center}
		\subfloat[\label{c_1}]{%
			\includegraphics[width=0.42\linewidth]{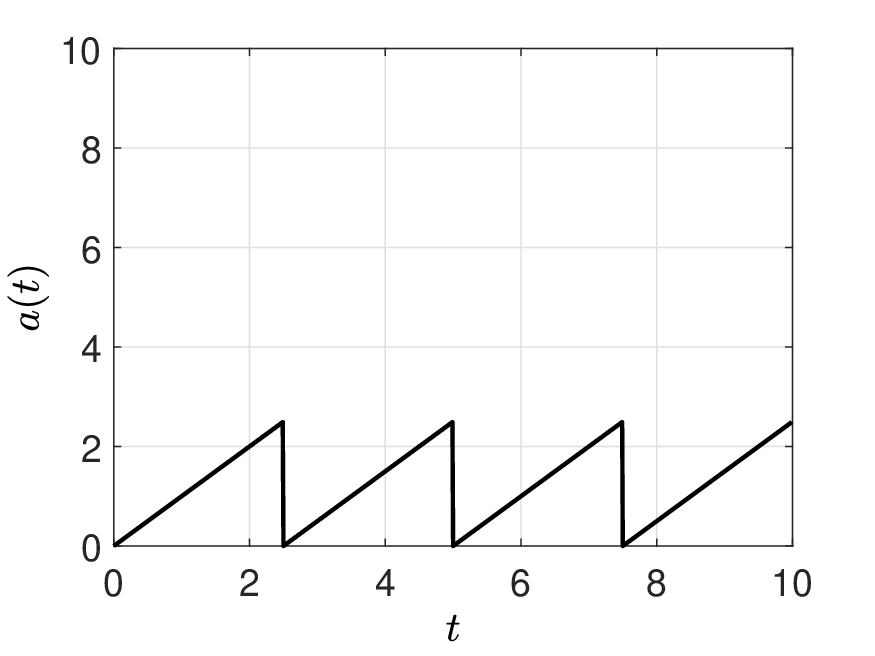}}\hfil
		\subfloat[\label{c_2}]{%
			\includegraphics[width=0.42\linewidth]{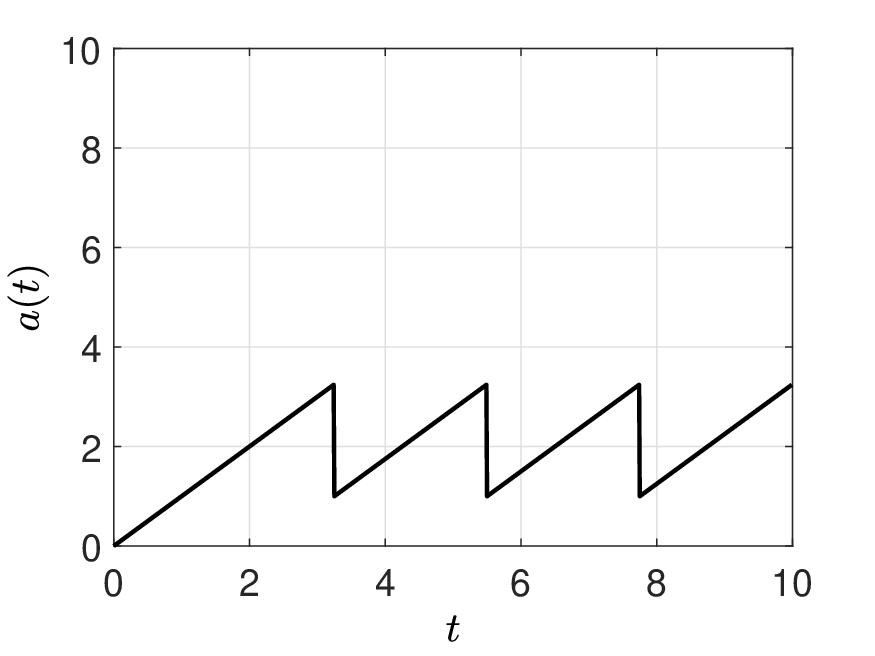}} \\
		\subfloat[\label{c_3}]{%
			\includegraphics[width=0.42\linewidth]{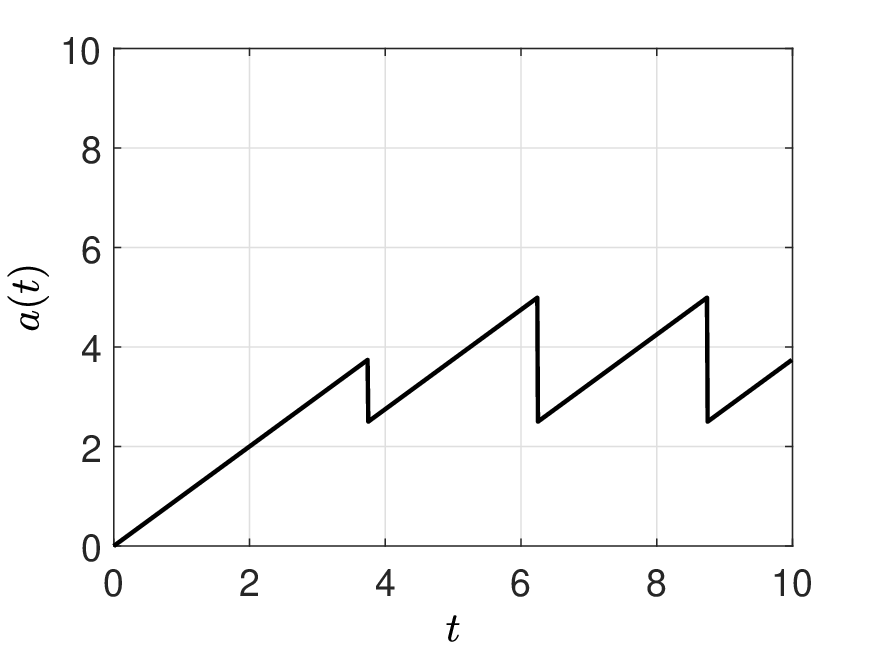}}\hfil
		\subfloat[\label{c4}]{%
			\includegraphics[width=0.42\linewidth]{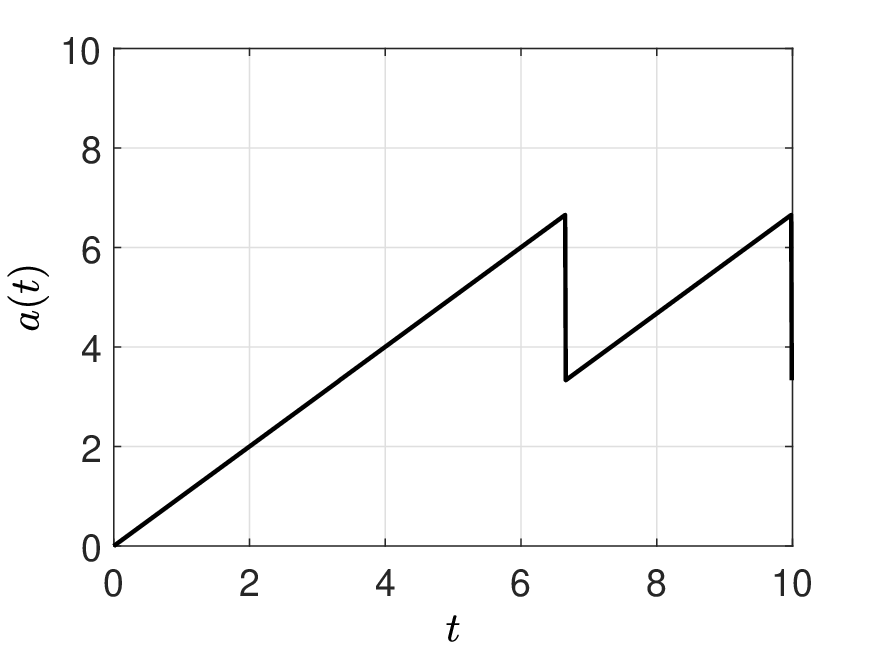}}
		\caption{Evolution of $a(t)$ with optimal update policies for $T =10$, $N=3$, (a) $c=0$, (b) $c=1$, (c) $c=2.5$, (d) $c=\frac{10}{3}$, when the maximum allowed distortion function is a constant.}
		\label{Fig:Sim_result}
	\end{center}
	\vspace{-4mm}
\end{figure*}

\subsection{Simulation Results for Constant Allowable Distortion }\label{subsect:constant}
 
We provide five numerical results for an exponentially decaying distortion function, $D_e$, defined in (\ref{dist_fnc}) with $a= (1-e^{-1})^{-1}$, $b=\frac{1}{4}$ and $d= e^{-1}$. Note that we can choose the processing time $c_i$ in $[0,4]$. When the processing time $c_i$ is equal to $0$, the distortion function $D_e(c_i)$ attains its maximum value, i.e., $D_e(c_i)= 1$. When the processing time $c_i$ is equal to $4$, the distortion function $D_e(c_i)$ reaches its minimum value, i.e., $D_e(c_i)= 0$. Since the maximum allowed distortion is a constant, we can rewrite the distortion constraint, $D_e(c_i)\leq \beta$, as $c_i\geq c$ where $c=D_e^{-1}(\beta)$ is in $[0,4]$. For the first four simulations, we cover each optimal policy given in Section~\ref{sub_sect:constant}. In these simulations, we take $T =10$ and $N=3$.

In the first example, we take $c=0$. In other words, there is no distortion constraint on the updates. In this case, the optimal policy is to request an update in equal time periods, i.e., $y_i =2.5$ for all $i$. As there is no distortion constraint on the updates, the information provider sends the updates immediately, i.e., $c_i=0$ for all $i$, and the updates have the highest possible distortion. As a result, the optimal age evolves as in Fig.~\ref{Fig:Sim_result}(a).

In the second example, we take $c=1$. This is the case where the minimum required processing time $c$ is small compared to the total time duration $T$, i.e., $(N+2)c<T$. In the optimal policy, the receiver waits for an equal amount of time to request another update after the previous update is received except a longer waiting time for the first update. The optimal age evolution is given in Fig.~\ref{Fig:Sim_result}(b). We note that the optimal policy is to request the first update after $s_1=2.25$ time. For the remaining updates, after the previous update is received, the receiver waits for $s_{2}=s_3=1.25$ time to request another update. After receiving a request, the provider generates the updates after processing $c=1$ time.

For the third example, we take $c=2.5$. In this case, the minimum required processing time is high which means that we wish to receive the updates with lower distortion compared to previous cases. The optimal age evolution is shown in Fig.~\ref{Fig:Sim_result}(c). We note that the optimal policy is to request the first update after waiting $s_1 = 1.25$. The receiver requests the remaining updates as soon as the previous update is received (back-to-back) since the provider uses relatively large amount of time to generate updates. In this case, the provider processes each update for $c_i = 2.5$ time for all $i$. 

\begin{figure}[t]
	\centerline{\includegraphics[width=0.95\columnwidth]{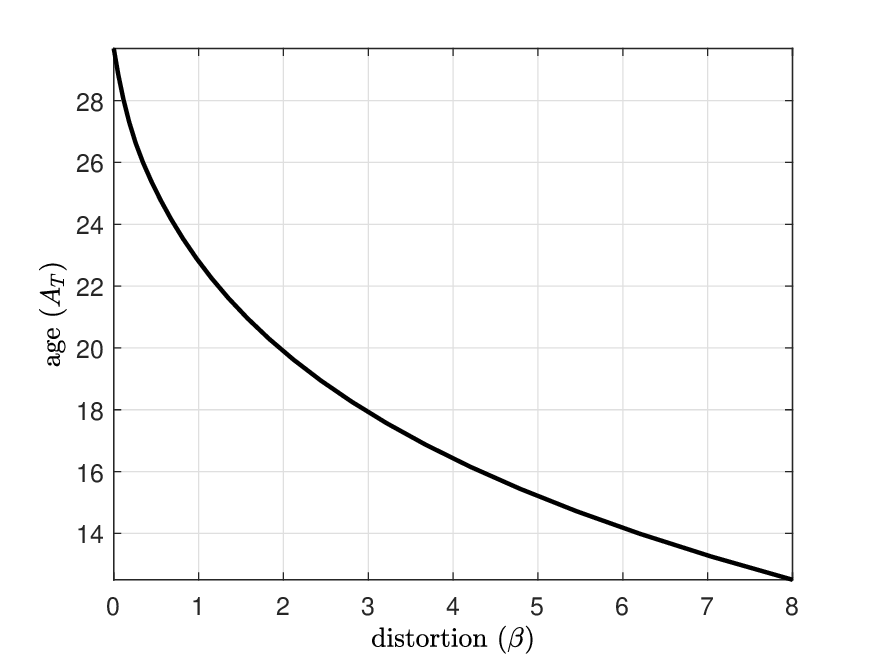}}
	\caption{Age versus distortion of the updates for $a =\frac{8}{1-e^{-3}}$, $b=1.2$, and $d =e^{-3}$ in (\ref{dist_fnc}) when the maximum allowed distortion is a constant. We vary $\beta$ and find the minimum age for each $\beta$.}
	\label{Fig:trade_off}
	\vspace{-4mm}
\end{figure}

For the fourth example, we take $c=\frac{10}{3}$ which is the highest possible minimum required processing time as $Nc=T$. In this case, there is only one feasible solution, which is to request the first update at $t=0$ and the remaining updates as soon as the previous update is received (back-to-back), i.e., $s_i= 0$ for all $i$. The provider processes each update for $c_i=\frac{10}{3}$ time for all $i$. The optimal age evolves as in Fig.~\ref{Fig:Sim_result}(d).  

Finally, we note that there is a trade-off between age and distortion. If we increase the distortion constraint $\beta$ (hence decrease the processing time constraint $c$), then we achieve a lower average age at the receiver, but the receiver obtains updates with low quality as the distortion of the updates is high. On the other hand, if we decrease the distortion constraint $\beta$ (hence increase the processing time constraint $c$), the receiver obtains updates with high quality, but in this case, the average age at the receiver increases. We show this trade-off between age and distortion as a fifth example in Fig.~\ref{Fig:trade_off}. 

Next, in the following subsection we provide numerical results for the case where the maximum allowed distortion function depends on the current age. 

\subsection{Simulation Results for Age-Dependent Allowable Distortion}

First, we provide two numerical results for the case where the maximum allowed distortion function is inversely proportional to the instantaneous age, i.e., we have $c_i\geq \alpha y_i$ constraint for each update. 

\begin{figure*}[t]
	\begin{center}
		\subfloat[]{%
			\includegraphics[width=0.45\linewidth]{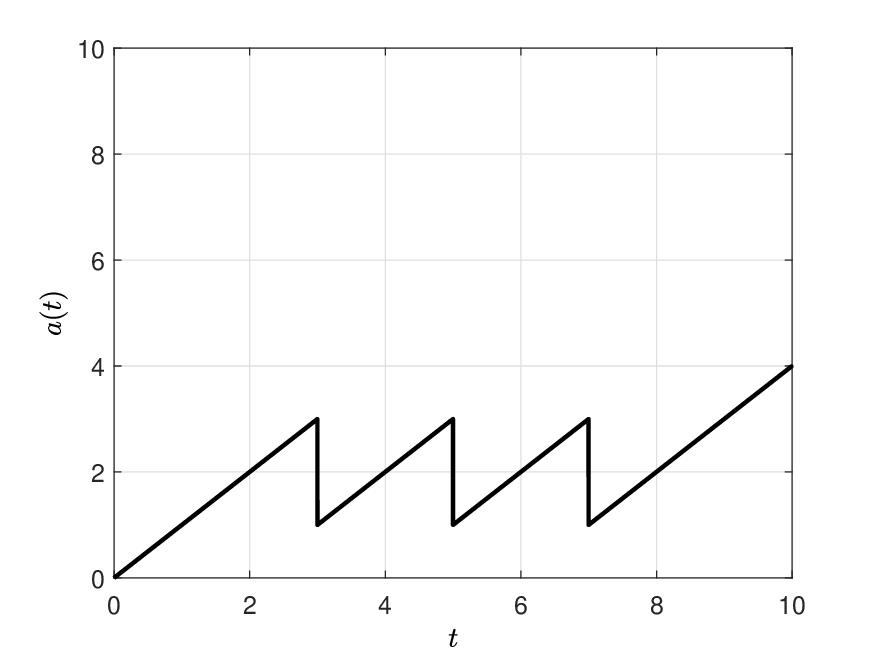}}\hfil
		\subfloat[]{%
			\includegraphics[width=0.45\linewidth]{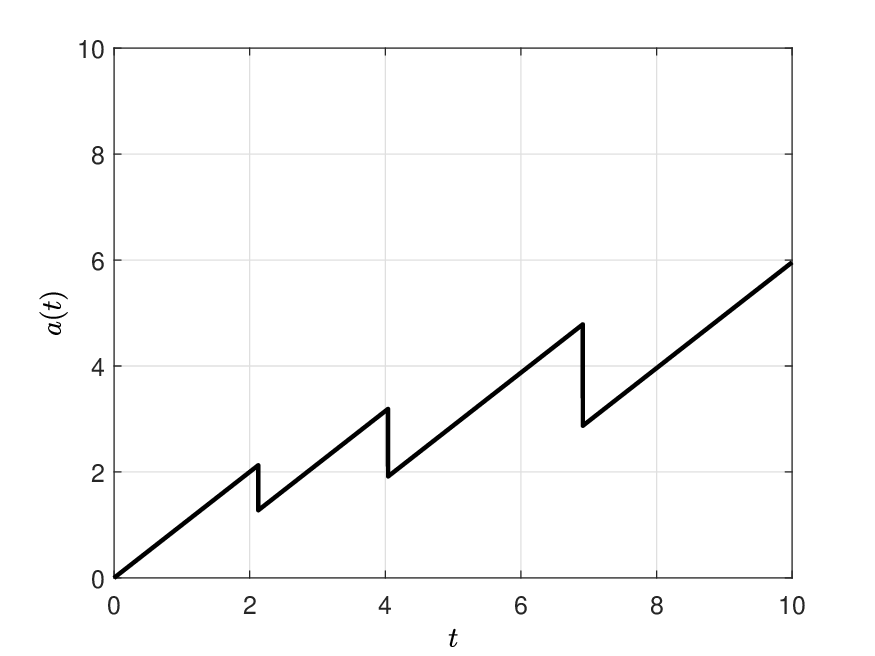}}
		\caption{Evolution of $a(t)$ with optimal update policies for $T =10$, $N=3$, (a) $\alpha=0.5$, (b) $\alpha=1.5$, when the maximum allowed distortion function is inversely proportional to the current age, i.e., $c_i\geq \alpha y_i$.}
		\label{Fig:Sim_result2}
	\end{center}
	\vspace{-4mm}
\end{figure*}

For the first example, we take $T=10$, $N=3$ and $\alpha =0.5$. This example corresponds to the case where the maximum allowed distortion slowly decreases with the current age, i.e., $\alpha$ is small. The optimal solution follows (\ref{opt_yi}) and (\ref{opt_y_N+1}) and is equal to $y_i =2$ for $i =1,2,3$ and $y_4 = 4$. We note that the information receiver requests all the updates when its age is equal to $y_i = 2$, and then, lets its age grow for the remaining time. Since $c_i = \alpha y_i$, we have $c_i = 1$ for all $i$ which means that all the updates have the same level of distortion as the processing times for the updates are equal. We observe in Fig.~\ref{Fig:Sim_result2}(a) that the optimal policy resembles the optimal policy for the case with constant allowable distortion when the minimum required processing time is small, i.e., the second example shown in Fig.~\ref{Fig:Sim_result}(b) in Section~\ref{subsect:constant}. 

For the second example, we take $T=10$, $N=3$ and $\alpha=1.5$. This example corresponds to the case where the maximum allowed distortion decreases faster with the instantaneous age, i.e., $\alpha$ is large. The optimal policy follows (\ref{age_opt_y1})-(\ref{age_opt_y_son}) and the optimal age evolution is shown in Fig.~\ref{Fig:Sim_result2}(b). The optimal solution is $y_1 = 0.8511$, $y_2= 1.2766$, $y_3 = 1.9149$ and $y_4 = 5.9574$. Due to $c_i = \alpha y_i$, we have $c_1 = 1.2766$, $c_2 = 1.9149$ and $c_3 = 2.8723$. We observe different from the first example where $\alpha<1$ that the processing time for each update is different which also means that updates have different levels of distortion. We also note that updates are requested right after the previous update is received except for the first update, i.e., $s_i =0$ for $i=2,\dots, N$.              

In the following four examples, we consider the case where the maximum allowed distortion function is proportional to the current age, i.e., we have $c_i\geq c-\alpha y_i$ constraint for each update. We take $N=3$, $c= 1$, $\alpha =0.4$. 

\begin{figure*}[t]
	\begin{center}
		\subfloat[]{%
			\includegraphics[width=0.45\linewidth]{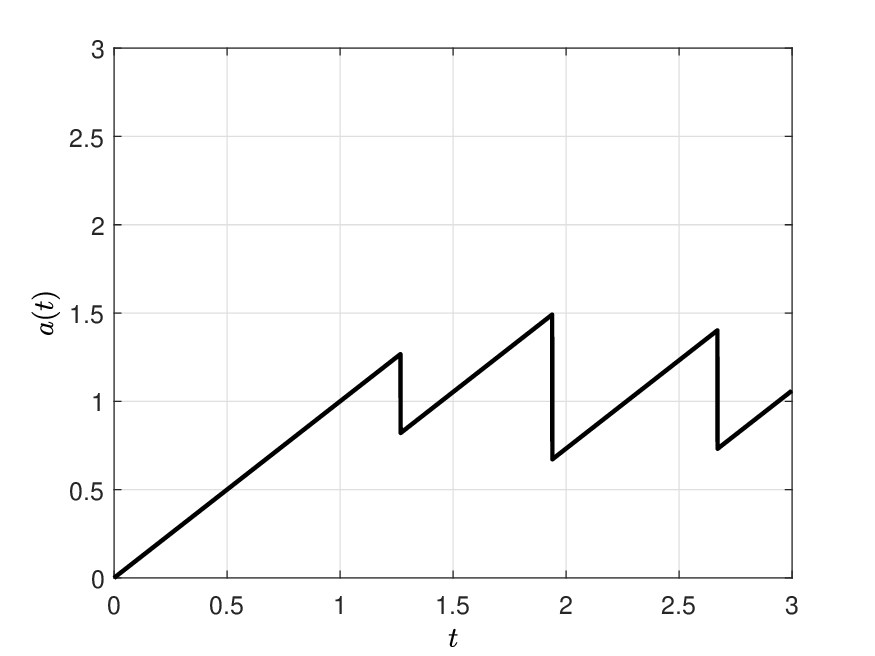}}\hfil
		\subfloat[]{%
			\includegraphics[width=0.45\linewidth]{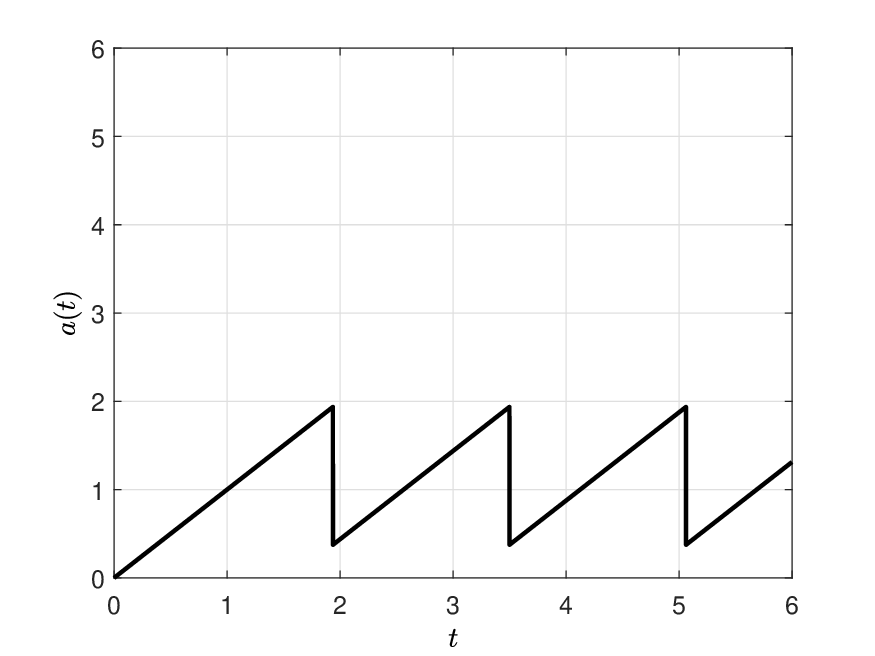}} \\
		\subfloat[]{%
			\includegraphics[width=0.45\linewidth]{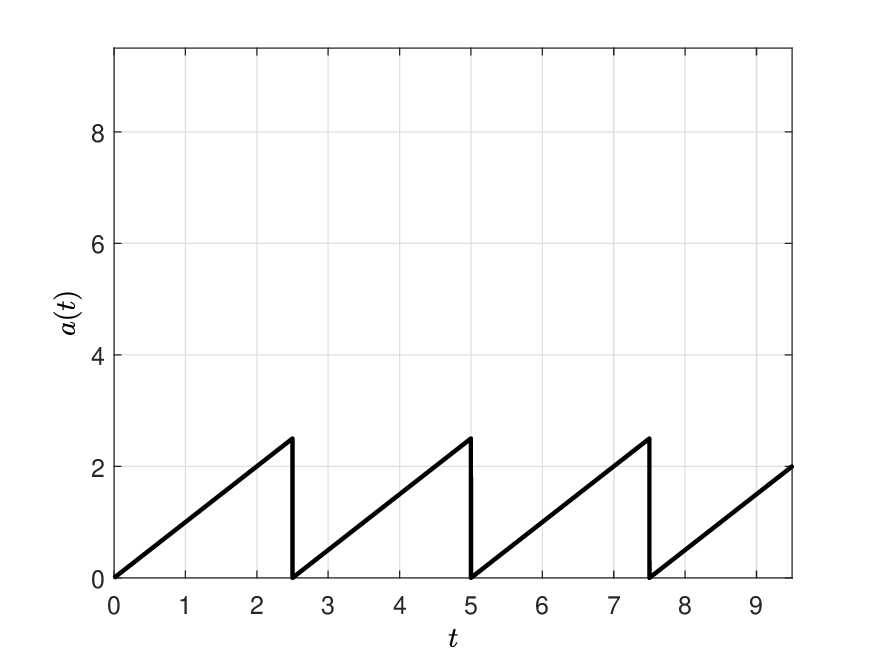}}\hfil
		\subfloat[]{%
			\includegraphics[width=0.45\linewidth]{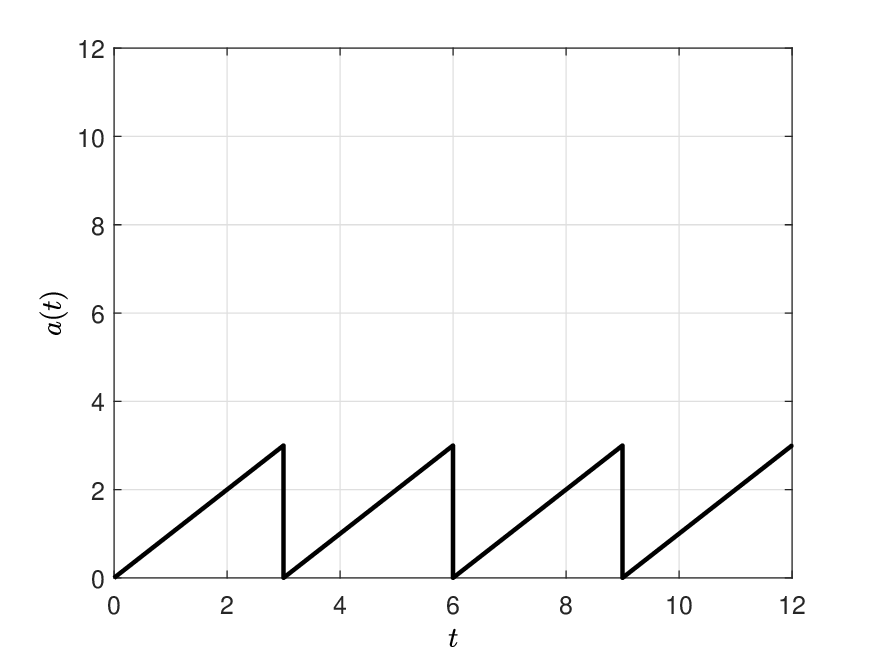}}
		\caption{Evolution of $a(t)$ with optimal update policies for $c =1$, $N=3$, $\alpha = 0.4$, (a) $T=3$, (b) $T=6$, (c) $T=9.5$, (d) $T=12$, when the maximum allowed distortion function is an increasing function of the current age, i.e., $c_i\geq c- \alpha y_i$.}
		\label{Fig:Sim_result3}
	\end{center}
	\vspace{-4mm}
\end{figure*}

For the first example, we take $T= 3$ which corresponds to the case where $T$ is relatively small compared to the minimum required processing time. The optimal policy follows (\ref{y_1_opt})-(\ref{y_N+1_opt}). The optimal solution is to choose $y_1 = 0.4478$, $y_2 = 0.8209$, $y_3=0.6716$ and $y_4 = 1.0597$. Since $c_i = c-\alpha y_i$, we have $c_1 = 0.8209$, $c_2 = 0.6716$ and $c_3 = 0.7313$. The optimal age evolution is shown in Fig.~\ref{Fig:Sim_result3}(a). We observe that updates are requested right after the previous update is received except for the first update, i.e., $s_i =0$ for $i=2,\dots,N$. In this case, as the instantaneous age is relatively low when the update is requested, the information provider processes the updates further to generate updates with high quality.

For the second example, we take $T=6$ which corresponds to the case where $T$ is relatively large compared to the minimum required processing time. The optimal solution follows (\ref{y_1_opt_case_2})-(\ref{y_N+1_opt_case_2}) and $y_1=y_2=y_3=1.5625$ and $y_4 = 1.3125$. We have $c_i=0.3750$ for all $i$. The optimal age evolution is shown in Fig.~\ref{Fig:Sim_result3}(b). As the instantaneous age is higher when the updates are requested compared to the first example, the system imposes a low distortion constraint for each update. We observe that as the receiver requests all the updates when the age at the receiver is equal to $y_i =1.5625$ for $i = 1,2,3$, the distortion constraint for each update becomes the same.       

For the third example, we take $T=9.5$ which corresponds to the case where the optimal policy follows $y_i =\frac{c}{\alpha}$ and $y_{N+1} = T- \sum_{i=1}^{N} y_i$. The optimal solution is $y_i =2.5$ for $i=1,2,3$ and $y_4 = 2$. In this case, as the instantaneous age gets higher when the update is requested, freshness of the updates becomes more important than the quality of the updates. That is why in this case, there is no active distortion constraints on the updates, i.e., $c_i\geq 0$. Thus, the receiver sends the updates without any processing, i.e., $c_i =0$ for all $i$. The optimal age evolution is shown in Fig.~\ref{Fig:Sim_result3}(c). Since the processing time for each update is equal to zero, the updates are not aged during the processing time and the age of the receiver reduces to zero after receiving each update. 

For the fourth example, we take $T=12$. The optimal policy follows $ y_i = \frac{T}{N+1}$ and is equal to $y_i = 3$ and $c_i = 0$ for all $i$. The optimal age evolution is shown in Fig.~\ref{Fig:Sim_result3}(d). In this case, we observe a similar optimal solution structure as in the previous case where $T=9.5$. As the updates are requested when the age is too high, updates with the highest distortion become acceptable for the system. We thus observe the same optimal solution structure as in the case with constant allowable distortion when there is no active distortion constraint, i.e., when $c=0$ in the first example shown in Fig.~\ref{Fig:Sim_result}(a) in Section~\ref{subsect:constant}.       

\section{Conclusions and Discussion} \label{sect:rest_upt}

In this paper, we considered the concept of status updating with update packets subject to distortion. In this model, updates are generated at the information provider (transmitter) following an update generation process that involves collecting data and performing computations. The distortion in each update decreases with the processing time during update generation at the transmitter; while processing longer generates a better-precision update, the long processing time increases the age of information. This implies that there is a trade-off between precision (quality) of information and age (freshness) of information. The system may be designed to strike a desired balance between quality and freshness of information. In this paper, we determined this design, by solving for the optimum update scheme subject to a desired distortion level. 

We considered the case where the maximum allowed distortion does not depend on the current age, i.e., is a constant, and the case where the maximum allowed distortion depends on the current age. For this case, we considered two sub-cases, where the maximum allowed distortion is a decreasing function and an increasing function of the current age.

Finally, we note that while we formulated the allowable distortion constraint using the \emph{current age at the receiver}, we could similarly formulate it by using \emph{time elapsed since the last requested update}. Specifically, we could use the constraint $c_i\geq \alpha s_i$ instead of the constraint $c_i \geq \alpha y_i$ in (\ref{problem4}) and the constraint $c_i\geq c-\alpha s_i$ instead of the constraint $c_i\geq c-\alpha y_i$ in (\ref{problem6}). We note that these two considerations are similar: If the receiver has not requested an update for a long time (large $s_i$), its current age will be high (large $y_i$). Due to space limitations and in order to avoid repetitive arguments, in this paper, we only considered the case where the distortion constraint depends on the instantaneous age $y_i$ at the receiver at the time of requesting a new update.
	
\bibliographystyle{unsrt}
\bibliography{IEEEabrv,lib_v1_melih}
\end{document}